\def\year{2019}\relax
\documentclass[letterpaper]{article} 
\usepackage{aaai19}  
\usepackage{times}  
\usepackage{helvet} 
\usepackage{courier}  
\usepackage[hyphens]{url}  
\usepackage{graphicx} 
\def\UrlFont{\rm}  
\usepackage{graphicx}  
\usepackage{colortbl}
\usepackage{booktabs} 
\usepackage{footmisc}
\usepackage{multirow}
\usepackage{subcaption}
\usepackage{todonotes}

\nocopyright

\title{Thou Shalt Not Hate: Countering Online Hate Speech\thanks{Accepted at ICWSM 2019}}
\author{Binny Mathew, Punyajoy Saha,\textsuperscript{\rm 1} Hardik Tharad, Subham Rajgaria, Prajwal Singhania,\\ \Large \textbf{Suman Kalyan Maity,\textsuperscript{\rm 2} Pawan Goyal, Animesh Mukherjee}\\
Indian Institute of Technology(IIT), Kharagpur\\
\textsuperscript{\rm 1}Indian Institute of Engineering Science and Technology (IIEST), Shibpur\\
\textsuperscript{\rm 2}Kellogg School of Management, Northwestern University\\
binnymathew@iitkgp.ac.in, \{punyajoysaha1998, hardik.tharad, subham.rajgaria, prajwal1210\}@gmail.com,\\ suman.maity@kellogg.northwestern.edu, \{pawang, animeshm\}@cse.iitkgp.ac.in
}

\begin{document}

\maketitle

\begin{abstract}
Hate content in social media is ever increasing. While Facebook, Twitter, Google have attempted to take several steps to tackle the hateful content, they have mostly been unsuccessful. Counterspeech is seen as an effective way of tackling the online hate without any harm to the freedom of speech. Thus, an alternative strategy for these platforms could be to promote counterspeech as a defense against hate content. However, in order to have a successful promotion of such counterspeech, one has to have a deep understanding of its dynamics in the online world. Lack of carefully curated data largely inhibits such understanding. In this paper, we create and release the first ever dataset for counterspeech using comments from YouTube. The data contains 13,924 manually annotated comments where the labels indicate whether a comment is a counterspeech or not. This data allows us to perform a rigorous measurement study characterizing the linguistic structure of counterspeech for the first time. This analysis results in various interesting insights such as: the counterspeech comments receive much more likes as compared to the non-counterspeech comments, for certain communities majority of the non-counterspeech comments tend to be hate speech, the different types of counterspeech are not all equally effective and the language choice of users posting counterspeech is largely different from those posting non-counterspeech as revealed by a detailed psycholinguistic analysis. Finally, we build a set of machine learning models that are able to automatically detect counterspeech in YouTube videos with an F1-score of 0.71. We also build multilabel models that can detect different types of counterspeech in a comment with an F1-score of 0.60. 
\end{abstract}

\section{Introduction}

\begin{quote}
``If there be time to expose through discussion the falsehood and fallacies, to avert the evil by the processes of education, the remedy to be applied is more speech, not enforced silence.'' -- Louis Brandeis  
\end{quote}

The advent of social media has brought several changes to our society. It allowed people to share their knowledge and opinions to a huge mass in a very short amount of time. While the social media sites have been very helpful, they have some unintended negative consequences as well. One such major issue is the proliferation of hate speech~\cite{massaro1990equality}. To tackle this problem, several countries have created laws against hate speech\footnote{Hate speech Laws: \url{https://goo.gl/tALXsH}}. Organizations such as Facebook, Twitter, and YouTube have come together and agreed to fight hate speech as well\footnote{\url{https://goo.gl/sH87W2}}.

\subsection{Current protocols to combat hate speech and their limitations} 
One of the main tools that these organizations use to combat online hate speech is blocking or suspending the message or the user account itself. Although, several social media sites have taken strict actions to prohibit hate speech on websites they own and operate, they have not been very effective in this enterprise\footnote{\url{https://goo.gl/G7hNtS}, \url{https://goo.gl/zEu4aX}, \url{https://goo.gl/CFmsqM}}. At the same time, some have argued that one should not block/suspend free speech because selective free speech is a dangerous precedent. 

While blocking of hateful speech may reduce its impact on the society, one always has the risk of violation of free speech. Therefore, the preferred remedy to hate speech would be to add more speech ~\cite{richards2000counterspeech}.

\subsection{Can countering hate speech be an effective solution?}
This requirement led countries and organizations to consider countering of hate speech as an alternative to blocking~\cite{gagliardone2015countering}. The idea that `more speech' is a remedy for harmful speech has been familiar in liberal democratic thought at least since the U.S. Supreme Court Justice Louis Brandeis declared it in 1927. There are several initiatives with the aim of using counterspeech to tackle hate speech. For example, the Council of Europe supports an initiative called `No hate speech movement'\footnote{\label{nohate speechmovement}No hate speech Movement Campaign:  \url{http://www.nohate speechmovement.org/}} with the aim to reduce the levels of acceptance of hate speech and develop online youth participation and citizenship, including in Internet governance processes. UNESCO released a study~\cite{gagliardone2015countering} titled `Countering Online Hate Speech', to help countries deal with this problem. Social platforms like Facebook have started counterspeech programs to tackle hate speech\footnote{\label{counterfb} Counterspeech Campaign by Facebook: \url{https://counterspeech.fb.com/en/}}. Facebook has even publicly stated that it believes counterspeech is not only potentially more effective, but also more likely to succeed in the long run ~\cite{bartlett2015counter}. Combating hate speech in this way has some advantages: it is faster, more flexible and responsive, capable of dealing with extremism from anywhere and in any language and it does not form a barrier against the principle of free and open public space for debate. 

\subsection{Working definition of counterspeech} 
In this paper, we define counterspeech as a direct response/comment (not reply to a comment) that \textit{counters} the hateful or harmful speech. Taking the YouTube videos that contain hateful content toward three target communities: {\sl Jews}, {\sl African-American} (\textit{Blacks}) and {\sl LGBT}, we collect user comments to create a dataset which contains counterspeech. To annotate this dataset, we use the different classes of counterspeech described in \citeauthor{susan2016counterspeech}~\shortcite{susan2016counterspeech} with a slight modification to the `Tone' category. While the paper includes all kinds of tones in this category, we split this class further into two categories: `Positive tone' and `Hostile language'. Note that when we say that a comment is a counterspeech to a video, we are focusing on the person about whom the video is about. The video may contain other people as well (such as an interviewer).

\subsection{Our contributions and observations}
We annotate and release the first ever dataset\footnote{\label{dataset_link} \textbf{The dataset and models are available here:} \url{https://github.com/binny-mathew/Countering_Hate_Speech_ICWSM2019}} on counterspeech. The dataset is based on counterspeech targeted to three different communities: {\sl Jews}, {\sl Blacks}, and {\sl LGBT}. It consists of 6,898 comments annotated as counterspeech and an additional 7,026 comments tagged as non-counterspeech. The counterspeech comments are further labeled into one or more of the categories listed in Table~\ref{tab:dataset_statistics}.

While developing the dataset, we had several interesting observations. We find that overall counterspeech comments receive much more likes than non-counterspeech comments. Psycholinguistic analysis reveals striking differences between the language used by the users posting counter and non-counterspeech. We also observe that the different communities attract different proportions of counterspeech. `Humor' as a counterspeech seems to be more prevalent when {\sl LGBT} is the target community, while in case of the {\sl Jews} community, `Positive tone' of speech seems to be more widely used.

As an additional contribution, we define three classification tasks for the dataset and develop machine learning models: (a) counterspeech vs non-counterspeech classification, in which XGBoost performs the best with an F1-score of 0.71, (b) multi-label classification of the types of counterspeech present in a given counterspeech text, in which XGBoost performs the best, (c) cross-community classification with an F1-score in the range 0.62 - 0.65. With these tasks and analysis, we hope that our research can help in reducing the spread of hate speech online.

\section{Related work}

In this section, we review some of the related literature. ``Counter-speech is a common, crowd-sourced response to extremism or hateful content. Extreme posts are often met with disagreement/conflicts, derision, counter campaigns''~\cite{bartlett2015counter,Maity:2018}. \citeauthor{citron2011intermediaries}~\shortcite{citron2011intermediaries} categorizes four ways in which one can respond to hateful messages -- (i) \textbf{Inaction:} By not responding to the hate speech, we might be actually causing more harm. It sends a message that people do not care about the target community. (ii) \textbf{Deletion/Suspension:} The removal of hate speech is the most powerful option available in response to hate speech. Removal of the hateful content is sometimes accompanied by the removal or suspension of the user account as well. This strategy is used by most of the social networks such as Facebook, Twitter, Quora, etc. (iii) \textbf{Education:} Institutions can help in educating the public about hate speech and its implications, consequences and how to respond. Programmes such as `NO HATE SPEECH' movement\footref{nohate speechmovement} and Facebooks Counterspeech program\footref{counterfb} help in raising awareness, providing support and seeking creative solutions. (iv) \textbf{Counterspeech:} Counterspeech is considered as the preferred remedy to hate speech as it does not violate the normative of free speech. While government or organizations rarely take part in counterspeech, a large proportion of the counterspeech is actually generated by the online users.

Silence in response to digital hate carries significant expressive costs as well. When powerful intermediaries rebut demeaning
stereotypes (like the Michelle Obama image) and invidious falsehoods (such as holocaust denial), they send a powerful message to readers. Because intermediaries often enjoy respect and a sense of legitimacy, using counterspeech, they can demonstrate what it means to treat others with respect and dignity~\cite{citron2011intermediaries}.

While blocking might work as a counter at the individual scale, it might actually be detrimental for the community as a whole. Deletion of comments that seem hateful might affect a person's freedom of speech. Also, with blocking, it is not possible to recover from the damage that the message has already caused. Counterspeech can therefore be regarded as the most important remedy which is constitutionally preferred~\cite{benesch2014countering}.

Counterspeech has been studied on social media sites like Twitter~\cite{wright2017vectors,susan2016counterspeech}, YouTube~\cite{ernst2017hate} and Facebook~\cite{schieb2016governing}. \citeauthor{wright2017vectors} ~\shortcite{wright2017vectors} study the conversations on Twitter, and find that some arguments between strangers lead to favorable change in discourse and even in attitudes. \citeauthor{ernst2017hate}~\shortcite{ernst2017hate} study the comments in YouTube counterspeech videos related to Islam and find that they are dominated by messages that deal with devaluating prejudices and stereotypes corresponding to Muslims and/or Islam.
\citeauthor{schieb2016governing}~\shortcite{schieb2016governing} study counterspeech on Facebook and through simulation, find that the defining factors for the success of counterspeech are the proportion of the hate speech and the type of influence the counter speakers can exert on the undecided. \citeauthor{Stroud2018feminist}~\shortcite{Stroud2018feminist} perform case studies on feminist counterspeech. 
Another line of research considers ascertaining the success of the counterspeech. \citeauthor{susan2016successfullcounter}~\shortcite{susan2016successfullcounter} describes strategies that have favorable impact or are counterproductive on users who tweet hateful or inflammatory content. \citeauthor{munger2017tweetment}~\shortcite{munger2017tweetment} conducted an experiment to examine the impact of group norm promotion and social sanctioning on racist online harassment and found that subjects who were sanctioned by a high-follower in-group male significantly reduced their use of a racist slur.

Our work is different from the existing literature in several ways. As noted in ~\cite{susan2016counterspeech}, the literature for counterspeech is pretty less. The existing literature on counterspeech have been qualitative and anecdotal in nature, while ours is the first work which tries to study counterspeech empirically. ~\cite{wright2017vectors} noted that ``Computational approaches are required in order to study  and  engage  counterspeech  efforts  at  scale and there is no work which perform automatic detection of counterspeech". In this work we attempt for the first time various learning algorithms to detect counterspeech. Further, one of our main contributions is that we release the first ever dataset for counterspeech identification. Our paper not only does the classification for the counterspeech task; we take it a step forward and do multi-label classification for various types of counterspeech.

\section{Dataset}

YouTube is one of the key online platforms on the Internet with 1.5 billion logged-in users visiting the site every month\footnote{\url{http://goo.gl/eEqWAt}}. Many of these videos contain hate speech targeted toward various communities. In this paper, we focus on such hateful videos and scrape their comment section.

\subsection{Data collection from YouTube}
In order to gather a diverse dataset, we focus on three target communities: {\sl Jews}, {\sl Blacks}, and {\sl LGBT}. The first step in our work involved finding videos that contained hateful content. In order to find such videos, we searched YouTube for phrases such as `I hate Jews', `I hate blacks' etc. We then manually select videos\footref{dataset_link} that contain some act of hate against one of these communities. Next, we use the YouTube comment scraper\footnote{YouTube Comment Scrapper: \url{http://ytcomments.klostermann.ca/}} to collect all the comments from the selected videos. Each comment had fields such as the comment text, username, date, number of likes, etc.

\subsection{Dataset annotation}

Annotations were performed by a group of two PhD students working in the area of social computing and three undergraduate students in computer science with ages ranging between 21-30.

There are different types of counterspeech that have different effects on the user. In order to understand the differences between them, we annotate the dataset at two levels. 

\noindent\textbf{First level annotation:} In the first level, we select comments from the hate speech video and ask the annotators to annotate each of these comments as a counter/non-counter to the hate message/action in the video. We define a comment as counterspeech if it opposes the hatred expressed in the video. We only consider those comments which are direct response to the video and ignore all the replies to these comments as we observe that they usually tend to drift off-topic and the discussion becomes more personal and noisy. Each comment has been annotated by two users and the conflicting cases have been resolved by a third annotator. We achieve 90.23\% agreement between the two annotators with a Cohen's $\kappa$ of 0.804. As a result of this step, we arrive at 6,898 counterspeech comments and 7,026 non-counterspeech comments. To our surprise, we find that 49.5\%  of the direct responses to the selected hate videos are counterspeech.

\noindent\textbf{Second level annotation:} In order to obtain a deeper understanding of the types of counterspeech, we perform a second level annotation. We give the annotators a counterspeech text and ask them to label all the types of counterspeech that are present in it. We use the taxonomy of counterspeech described in \citeauthor{susan2016counterspeech}~\shortcite{susan2016counterspeech} for this purpose. For ease of readability we describe these categories in the subsequent section. 

Two independent annotators tagged each comment annotated as counterspeech in the first level into appropriate types. We obtain a loose $\kappa$ score of 0.868 and a strict $\kappa$ score of 0.743 for this task~\cite{ravenscroft2016multi}. We employ a third annotator for deciding on the conflicting cases. The final distribution of the different types of counterspeech is noted in Table~\ref{tab:dataset_statistics}.

\begin{table}[htb]
\centering
	\resizebox{0.95\columnwidth}{!}{\begin{tabular}{|l|c|c|c|c|c|} 
			\hline
			&\multicolumn{3}{|c|}{Target community}& Total\\
			\hline
			Type of counterspeech & {\sl Jews} & {\sl Blacks}  & {\sl LGBT} & \\
			\hline 
						Presenting facts & 308 & 85 & 359 & 752 \\
Pointing out hypocrisy or contradictions & 282 & 230 & 526 & 1038 \\
Warning of offline or online consequences & 112 & 417 & 199 & 728 \\
Affiliation & 206 & 159 & 200 & 565 \\
Denouncing hateful or dangerous speech & 376 & 482 & 473 & 1331 \\
Humor & 227 & 255 & 618 & 1100 \\
Positive tone & 359 & 237 & 268 & 864 \\
Hostile & 712 & 946 & 1083 & 2741 \\ \hline
Total & 2582 & 2811 & 3726 & 9119 \\
			
			\hline
		\end{tabular}}
		\caption{Statistics of the counterspeech dataset. Numbers corresponding to each of the target community, grouped as per the type of counterspeech are shown. Note that if a comment utilizes multiple strategies, we would include that particular comment in all the corresponding counterspeech types. Thus, we have a total of 9,119 counterspeech from 6,898 comments.}
		~\label{tab:dataset_statistics}
\end{table}

\subsection{Types of counterspeech}

\begin{figure}[h!]
	\centering
	\includegraphics[width=.95\columnwidth]{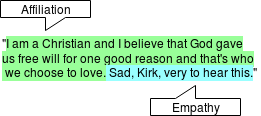}
	\caption{An example comment containing two types of counterspeech: \textit{affiliation} and \textit{empathy}. This comment was in response to a interview video in which the interviewee (Kirk) says that homosexuality is unnatural, detrimental and destructive to the society.}
	\label{fig:counter_multi_response}

\end{figure}

There are numerous strategies that could be used to counter the hateful messages in the online social media. \citeauthor{susan2016counterspeech}~\shortcite{susan2016counterspeech} distinguishes eight such strategies that are used by counterspeakers.
We decided on using these eight types of counterspeech with a slight modification to the category `Tone'. The authors have considered the whole spectrum of Tone as a single category. While one end of the spectrum (`Hostile') can cause the original speaker to delete his post/account and thus is unlikely to de-escalate the conversation, the other end of the spectrum (`Positive tone') could help in generating a positive attitude and thus de-escalate the conversation. So, we decided to divide the `Tone' category into  `Positive tone' and `Hostile language' categories for our work. 
Note that a single comment can consist of multiple types of counterspeech as shown in Figure~\ref{fig:counter_multi_response}. Also, the types of countespeech strategies discussed here are not comprehensive; there could be other types as well. In this paper, we focus on just these eight types of counterspeech. Below, we discuss these various categories.

\noindent\textbf{Presenting facts to correct misstatements or mis-perceptions:} In this strategy, the counterspeaker tries to persuade by correcting misstatements. An example of this type of counterspeech toward the {\sl LGBT} community from our dataset is as follows:~\textquotedblleft \textit{Actually homosexuality is natural. Nearly all known species of animal have their gay communities. Whether it be a lion or a whale, they have or had (if they are endangered) a gay community. Also marriage is an unnatural act. Although there are some species that do have longer relationships with a partner most known do not}\textquotedblright. This comment was in response to a interview video in which the interviewee says that homosexuality is unnatural, detrimental and destructive to the society.

\noindent\textbf{Pointing out hypocrisy or contradictions:}
In this strategy, the counterspeaker points out the hypocrisy or contradiction in the user's (hate) statements. In order to discredit the accusation, the 
individual may explain and rationalize their previous behavior, or if they are persuadable, resolve to avoid the dissonant behavior in the future~\cite{beauvois1993cognitive}. An example of this type of counterspeech toward {\sl LGBT} community from our dataset is as follows: \textquotedblleft \textit{The `US Pastor' can't accept gays because the Bible says not to be gay. But...he ignores:The thing about eating shrimp or pork, The thing about touching the skin of a dead pig (Football), The thing about mixing fabrics, The thing about torn clothes, The thing about tattoos, The thing about planting two crops in one field, The thing about working on the Holy Day (Saturday or Sunday depending)...for any and all of those sins one should burn for an eternity, yet is ignored. But when it comes to loving the wrong person (gays) this will not do! Christians only follow the parts of the bible that supports their bigotry. YOUR A HYPOCRITE.} \textquotedblright. This comment was in response to a interview video in which the interviewee encourages hate against homosexual people.

\noindent\textbf{Warning of offline or online consequences:}
In this strategy, counterspeaker warns the user of possible consequences of his/her actions. This can sometimes cause the original speaker of the hate speech to retract from his/her original opinion. An example of this type of counterspeech toward the {\sl LGBT} community from our dataset is as follows: \textquotedblleft \textit{I'm not gay but nevertheless, whether You are beating up someone gay or straight, it is still an assault and by all means, this preacher should be arrested for sexual harassment and instigating!!!} \textquotedblright. This comment was in response to a video in which a preacher advised people to beat  the kids if they are gay.

\noindent\textbf{Affiliation:}
Affiliation is \textquotedblleft...establishing, maintaining, or restoring a positive affective relationship with another person or group of persons\textquotedblright ~\cite{byrne1961anxiety}. People are more likely to credit the counterspeech of those with whom they affiliate, since they tend to ``evaluate ingroup members as more trustworthy, honest, loyal, cooperative, and valuable to the group than outgroup members'' ~\cite{kane2005knowledge}. In our dataset, couterspeakers who use {\sl Affiliation} receive the highest number of likes for their comments among all the counterspeech types. An example of this type of counterspeech toward the {\sl LGBT} community from our dataset is as follows:~\textquotedblleft \textit{Hey I'm Christian and I'm gay and this guy is so wrong. Stop the justification and start the accepting. I know who my heart and soul belong to and that's with God: creator of heaven and earth. We all live in his plane of consciousness so it's time we started accepting one another. That's all} \textquotedblright. This comment was in response to a interview video in which the interviewee encourages hate against homosexual people.

\noindent\textbf{Denouncing hateful or dangerous speech:}
In this strategy, the counterspeakers denounce the message as being hateful. This strategy can help the counterspeakers in reducing the impact of the hate message. An example of this type of counterspeech toward {\sl Jews} community from our dataset is as follows: \textquotedblleft \textit{please take this down YouTube. this is hate speech.} \textquotedblright. This comment was in response to a video in which a preacher is advocating hatred and killing of Jewish people.

\begin{figure*}[!ht]

\begin{subfigure}{.7\columnwidth}
  \centering
  \includegraphics[width=\columnwidth]{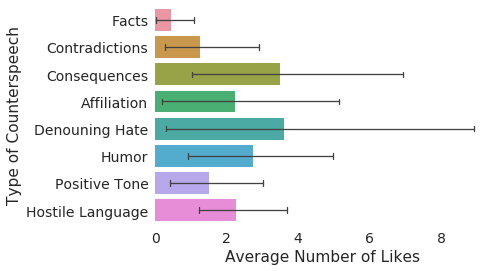}
  \caption{{\sl Blacks} community likes}
  \label{fig:sfig2}
\end{subfigure}%
\begin{subfigure}{.7\columnwidth}
  \centering
  \includegraphics[width=\columnwidth]{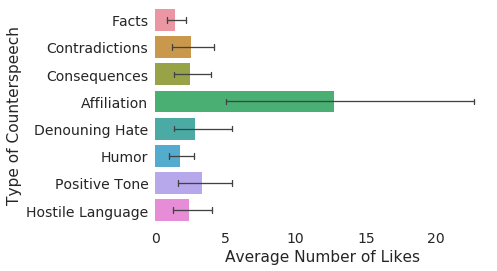}
  \caption{{\sl Jews} community likes}
  \label{fig:sfig3}
\end{subfigure}%
\begin{subfigure}{.7\columnwidth}
  \centering
  \includegraphics[width=\columnwidth]{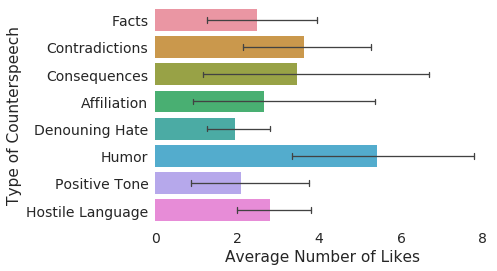}
  \caption{{\sl LGBT} community likes}
  \label{fig:sfig4}
\end{subfigure}%

\begin{subfigure}{.7\columnwidth}
  \centering
  \includegraphics[width=\columnwidth]{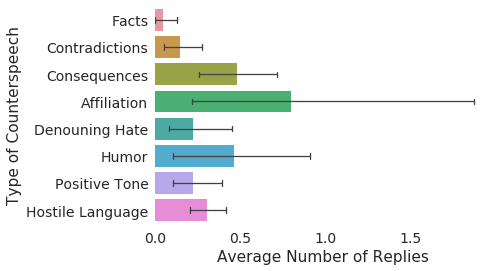}
  \caption{{\sl Blacks} community replies}
  \label{fig:sfig6}
\end{subfigure}%
\begin{subfigure}{.7\columnwidth}
  \centering
  \includegraphics[width=\columnwidth]{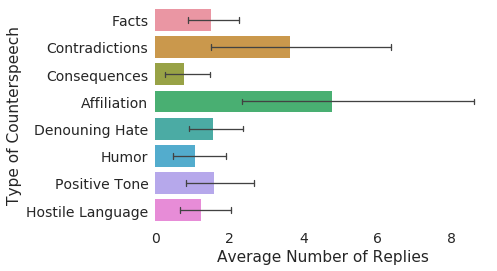}
  \caption{{\sl Jews} community replies}
  \label{fig:sfig7}
\end{subfigure}%
\begin{subfigure}{.7\columnwidth}
  \centering
  \includegraphics[width=\columnwidth]{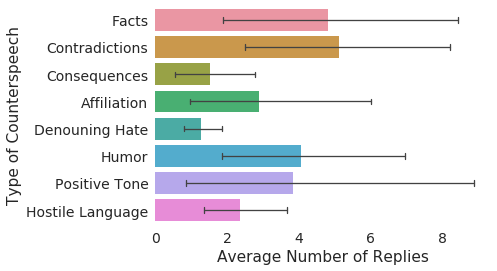}
  \caption{{\sl LGBT} community replies}
  \label{fig:sfig8}
\end{subfigure}%
\caption{Plots showing average number of likes and replies received by different types of counterspeech in the three communities. The whiskers represent 95\% confidence interval.}
\label{fig:avg_likes_comments}

\end{figure*}

\noindent\textbf{Humor and sarcasm:}
Humor is one of the most powerful tools used by the counterspeakers to combat hate speech. It can de-escalate conflicts and can be used to garner more attention toward the topic. Humor in online settings also eases hostility, offers support to other online speakers, and encourages social cohesion~\cite{marone2015online}. Often, the humor is sarcastic, like the following counterspeech comment subscribing the {\sl LGBT} community from our dataset:~\textquotedblleft \textit{HAHAHAHAHAHAHAH...oh you were serious. That's even funnier :)}\textquotedblright. This comment was in response to a video in which a preacher advocates hate and killing of homosexual people.

\noindent\textbf{Positive tone:}
The counterspeaker uses a wide variety of tones to respond to hate speech. In this strategy, we consider different forms of speech such as empathic, kind, polite, or civil. Increasing empathy with members of opposite groups counteracts
incitement~\cite{benesch2014countering}. We would like to point out that the original authors actually defined Tone to contain hostile counterspeech as well. Instead, we decide to make `Hostile language' as a separate type of counterspeech. An example of this type of counterspeech toward {\sl Jews} community from our dataset is as follows:~\textquotedblleft \textit{I am a Christian, and I believe we're to love everyone!! No matter age, race, religion, sex, size, disorder...whatever!! I LOVE PEOPLE!! We are not going to go anywhere as a country if we don't put God first in our lives, and treat EVERYONE with respect}\textquotedblright. This comment was in response to a video in which a preacher is advocating hatred and killing of Jewish people.

\noindent\textbf{Hostile language:}
In this strategy, the counterspeaker uses abusive, hostile, or obscene comments in response to the original hate message. Such a response can persuade an original speaker to delete his message or even a whole account, but is unlikely to either de-escalate the conversation or persuade the original speaker to recant or apologize. An example of this type of counterspeech toward {\sl African-American} from our dataset is as follows:~\textquotedblleft \textit{This is ridiculous!!!!!! I hate racist people!!!! Those police are a**holes!!!}\textquotedblright. This comment was in response to a video in which the police are performing a hate crime against.

\section{Detailed analysis}

In this section, we perform a detailed analysis over the dataset. We observe that 71.24\% of the counterspeech comments belong to exactly one counterspeech category. Thus, majority of the counterspeakers rely on a single strategy for counterspeech. As noted in Table~\ref{tab:dataset_statistics}, different communities attract different types of counterspeech. We observe that `Hostile language' is the major category among all the classes and is present in around 39.74\% of the counterspeech. Other than that, the counterspeakers for the {\sl Jews} community seem to be using `Positive tone' strategy in their counterspeech more often, while the counterspeakers of the {\sl LGBT} community more often use `Humor' and `Pointing out hypocrisy or contradiction' to tackle the hate speech.

\begin{figure*}[!ht]
\begin{subfigure}{.5\columnwidth}
  \centering
  \includegraphics[width=\columnwidth]{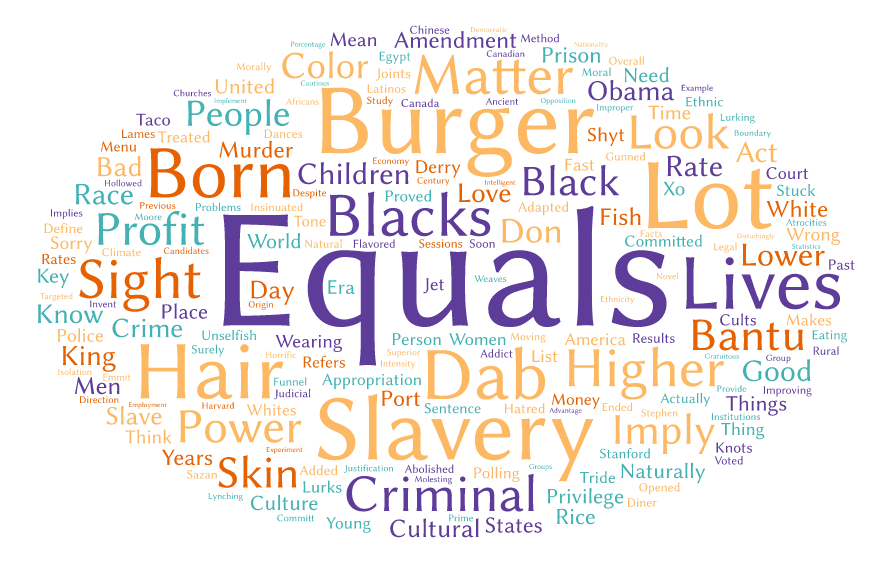}
  \caption{Presenting facts}
  \label{fig:wordcloud_category1_black}
\end{subfigure}
\begin{subfigure}{.5\columnwidth}
  \centering
  \includegraphics[width=\columnwidth]{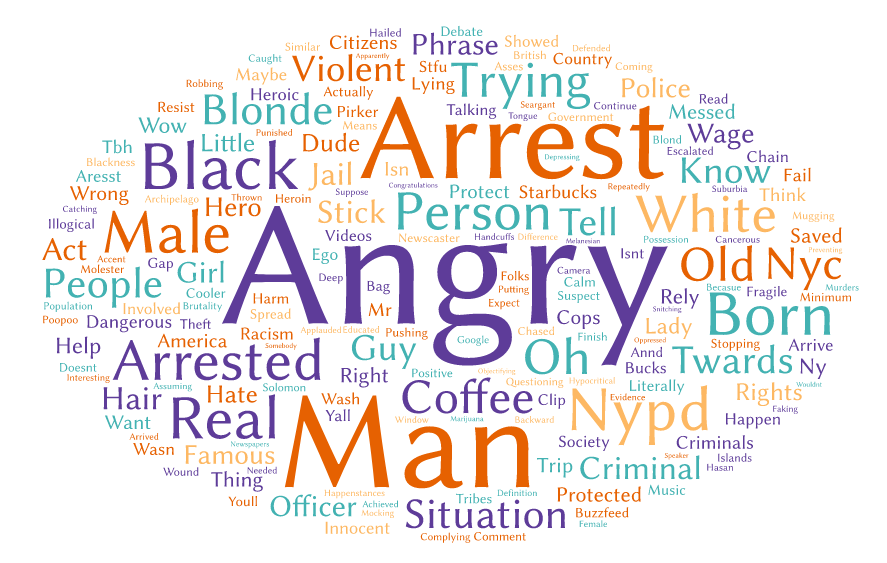}
  \caption{Pointing out hypocrisy}
  \label{fig:wordcloud_category2_black}
\end{subfigure}%
\begin{subfigure}{.5\columnwidth}
  \centering
  \includegraphics[width=\columnwidth]{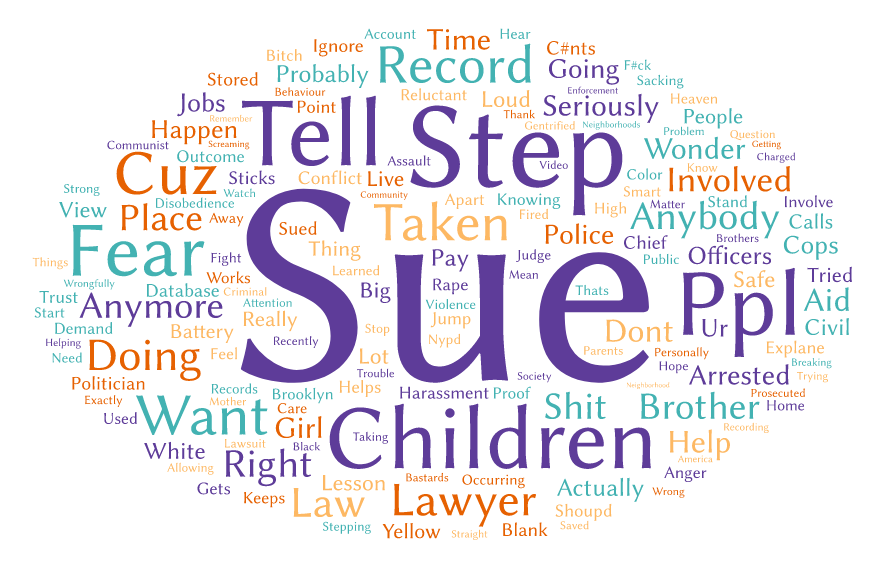}
  \caption{Warning of consequences}
  \label{fig:wordcloud_category3_black}
\end{subfigure}%
\begin{subfigure}{.5\columnwidth}
  \centering
  \includegraphics[width=\columnwidth]{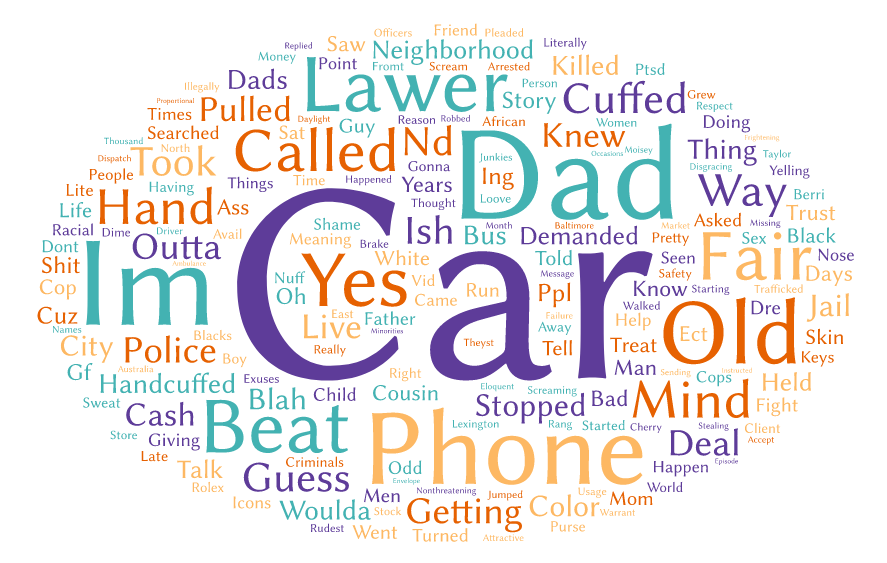}
  \caption{Affiliation}
  \label{fig:wordcloud_category4_black}
\end{subfigure}%

\begin{subfigure}{.5\columnwidth}
  \centering
  \includegraphics[width=\columnwidth]{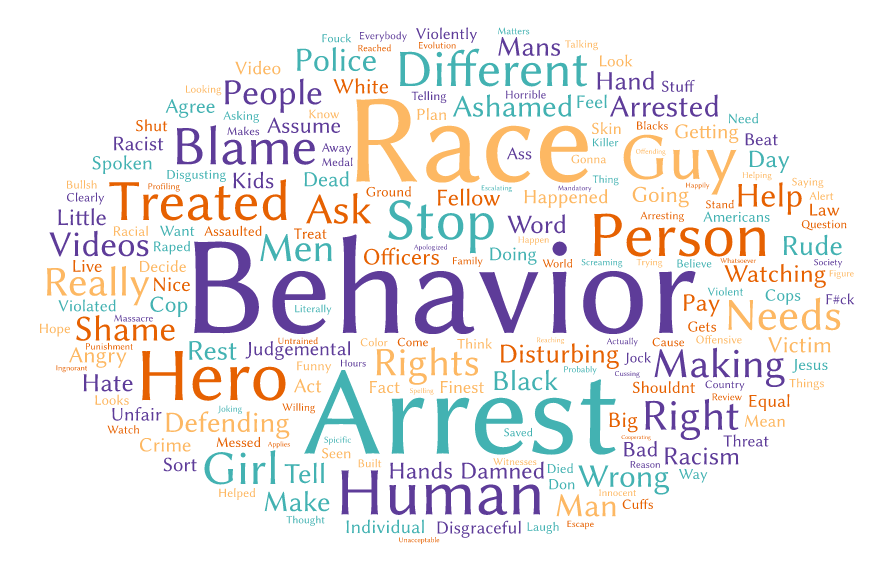}
  \caption{Denouncing speech}
  \label{fig:wordcloud_category5_black}
\end{subfigure}
\begin{subfigure}{.5\columnwidth}
  \centering
  \includegraphics[width=\columnwidth]{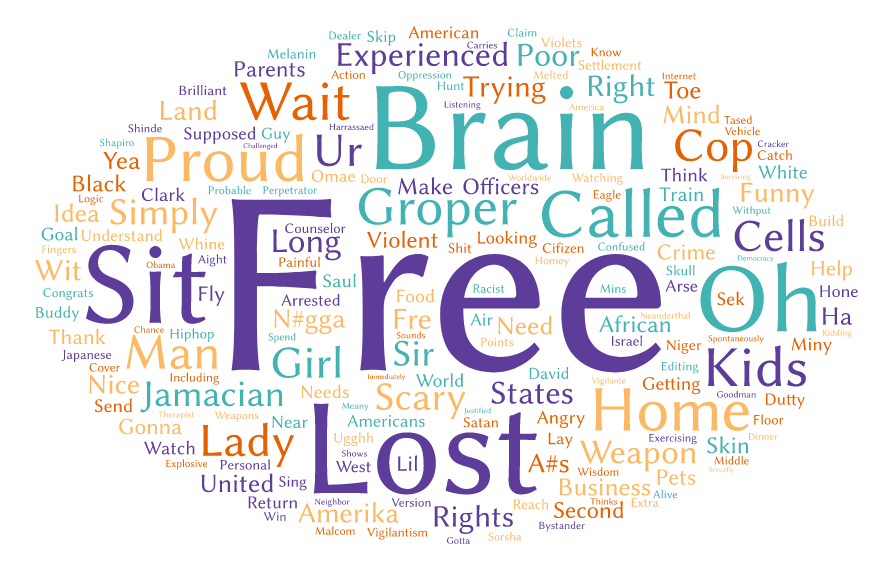}
  \caption{Humor and sarcasm}
  \label{fig:wordcloud_category6_black}
\end{subfigure}%
\begin{subfigure}{.5\columnwidth}
  \centering
  \includegraphics[width=\columnwidth]{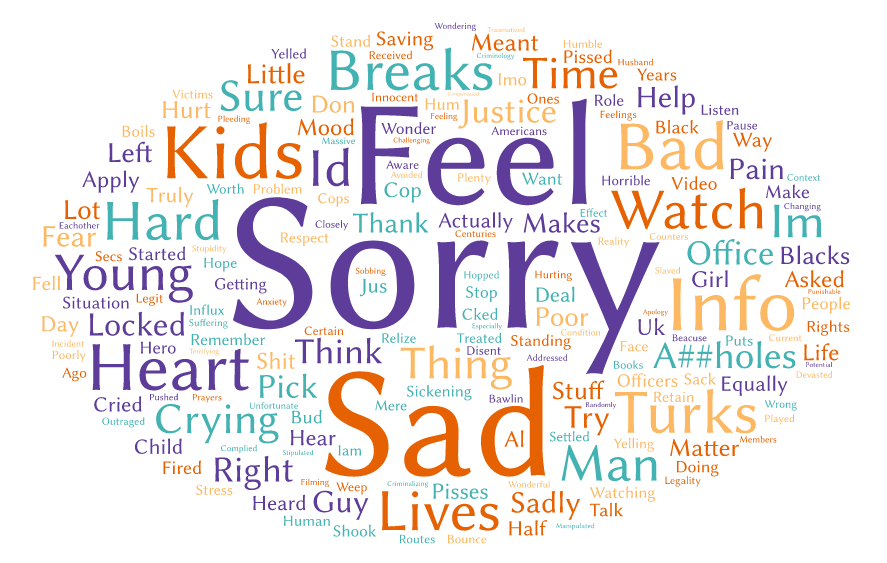}
  \caption{Positive tone}
  \label{fig:wordcloud_category7_black}
\end{subfigure}%
\begin{subfigure}{.5\columnwidth}
  \centering
  \includegraphics[width=\columnwidth]{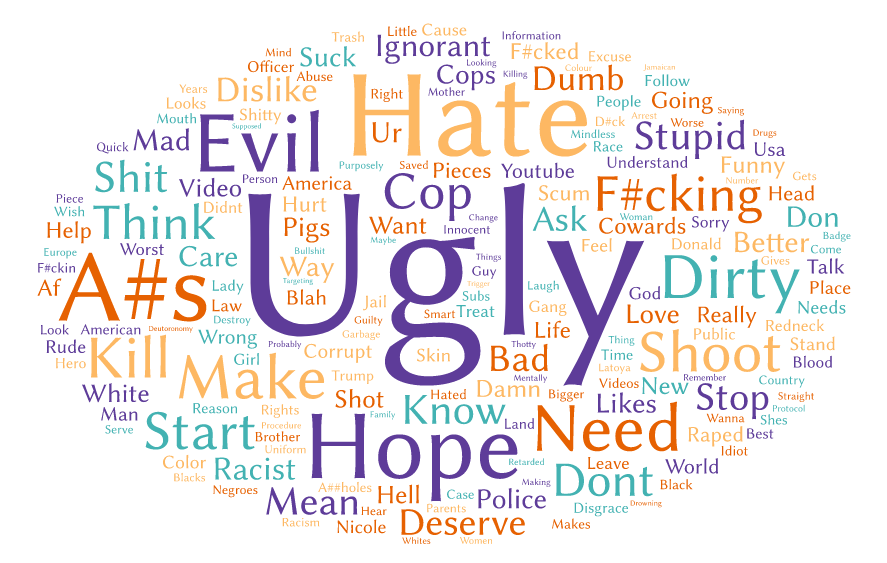}
  \caption{Hostile language}
  \label{fig:wordcloud_category8_black}
\end{subfigure}%
\caption{Word clouds for the different types of counterspeech used by the counterspeaker for hate speech against \textbf{Blacks}.}
\label{fig:wordcloud_counterspeech_type_black}

\end{figure*}

\subsection{Likes and comments}
We first analyze the comments as per the likes and comments received. We consider two groups - counterspeech comments and non-counterspeech comments. For our analysis, we also perform Mann–Whitney U test~\cite{mann1947test} to compare the two distributions.

On average, we find counterspeech comments in our data receiving 3.0 likes, in contrast to non-counterspeech comments receiving 1.73 likes, which is very less as compared to the counterspeech ($p\sim0.0$). Similarly, we investigate into the number of replies received and find that counterspeech comments receive more replies (average: $1.94$) than non-counterspeech comments (average: $1.50$). However, the differences were not as significant ($p>0.1$). 

We would also like to point out that the average likes and replies in YouTube videos are generally less. In ~\cite{thelwall2012commenting}, the authors analyze large  samples  of  text comments in YouTube videos. They found that 23.4\% of YouTube comments in complete comment  sets  are replies (called reply-density). The authors also analyzed the types of videos receiving the least and the most replies and found that videos pertaining to news, politics, and religion are at the top. This point is exemplified by our dataset where we found the reply density to be 45.17\%, which is almost double of a normal YouTube video. If we consider just the counterspeech comments, the reply density is even higher at 72.07\%. Due to the controversial nature of the content, it is expected to generate such discussion. We can thus state that our dataset is representative enough. In another work by ~\cite{siersdorfer2014analyzing}, the authors analyzed over 6 million comments from 67,290 videos and found the average rating to be 0.61. This is much less than the average 3.0 likes for our counterspeech comments. 

We view the likes received by the counterspeech comments as an endorsement by the community. In this sense, we can observe from Figure~\ref{fig:avg_likes_comments} that different communities seem to like different types of counterspeech. In case of the \textit{African-American} community, the average likes received by the counterspeech category which `Warn of offline/online consequences' and `Denouncing of hateful/dangerous speech' seem to be more as compared to the other types. In these counterspeech comments, the counterspeakers call out for racism and talk about how the person in focus could be sued for his actions. In case of the \textit{Jews} community, `Affiliation' seems to be the most endorsed form of counterspeech. In the comments, we observe that the people affiliate with both the target and the source community (`Muslims', `Christians') to counter the hate message. In some of the comments, the counterspeakers identify themselves as belonging to the same community as that of the hate speaker and claim that the hate message is unacceptable. Previous works have shown that these kinds of counterspeech are successful in changing the attitude of the hate speaker~\cite{berger2013matters,munger2017tweetment}. In case of the \textit{LGBT} community, we can observe that the community endorses several types of counterspeech with the `Humor' and `Pointing out hypocrisy or contradiction' receiving more average likes than others. In these comments, the counterspeakers make use of sarcasm and provide several points which contradict the statements expressed by the hate speaker.

\subsection{Lexical analysis}

To understand the language used in each type of counterspeech, we look into the words that are employed by the counterspeakers. We observe that some of the words such as \textit{Islam, Jew, People, Muslim, Black} were present in all the categories with high frequency. In order to filter out such words that are common in all categories, we use the tf-idf method. First, we generate a tf-idf matrix using the whole corpus. Then, for each type of counterspeech, we use the matrix to rank the words according to their tf-idf values.
We display the top 200 words present in each type of counterspeech. The size of a word is proportional to its tf-idf value. We observe from the figures (summarized below) that the words present are representative of the categories that they belong to.

In Figure~\ref{fig:wordcloud_counterspeech_type_black}, we plot the word cloud for the different types of counterspeech employed for the African-American community. Observe the presence of words such as `Illogical' in Figure~\ref{fig:wordcloud_category2_black}, `Sue', `Lawyer' in Figure~\ref{fig:wordcloud_category3_black}, `Im', `Friend' in Figure~\ref{fig:wordcloud_category4_black}, `Race', `Racist' in Figure~\ref{fig:wordcloud_category5_black}, `Oh', `Funny' in Figure~\ref{fig:wordcloud_category6_black}, `Sorry', `Feel', `Heart' in Figure~\ref{fig:wordcloud_category7_black}. 

\begin{figure*}[!ht]
\begin{subfigure}{.5\columnwidth}
  \centering
  \includegraphics[width=\columnwidth]{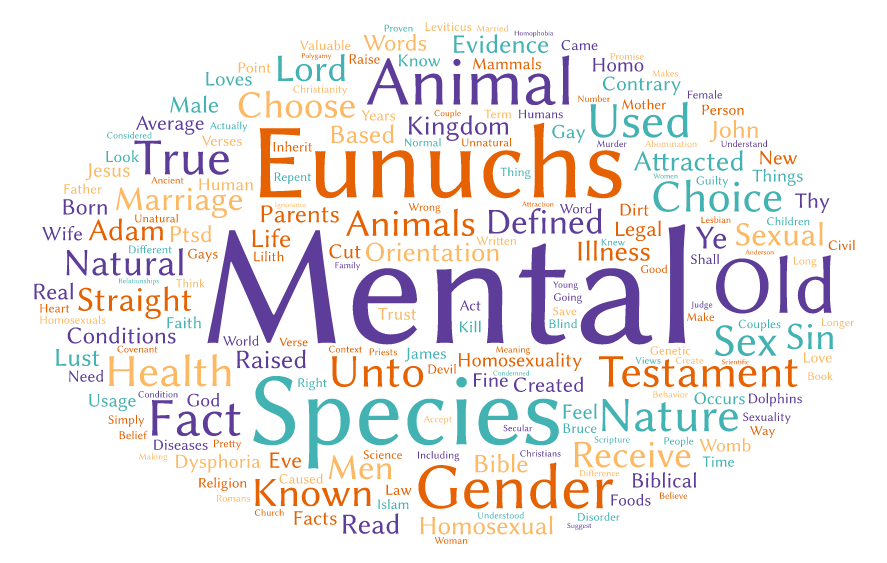}
  \caption{Presenting facts}
  \label{fig:wordcloud_category1_lgbtq}
\end{subfigure}
\begin{subfigure}{.5\columnwidth}
  \centering
  \includegraphics[width=\columnwidth]{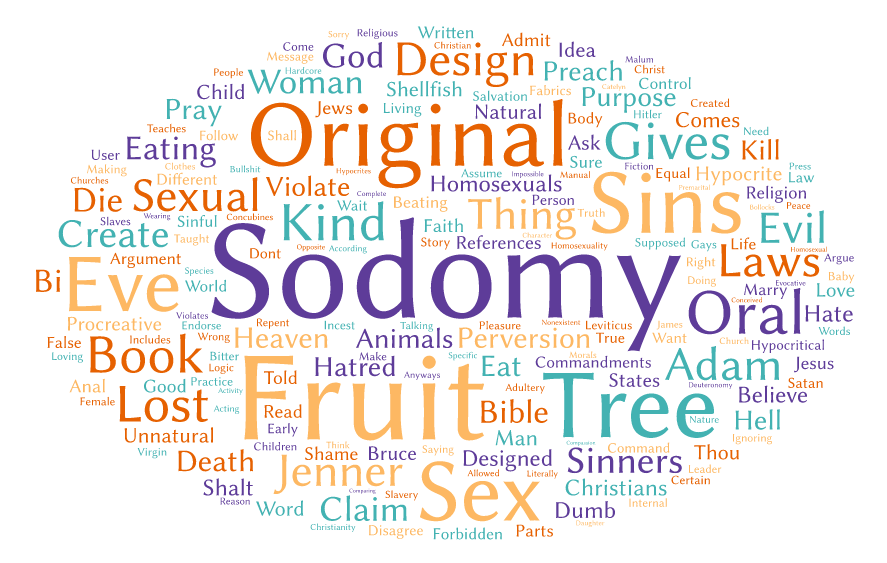}
  \caption{Pointing out hypocrisy}
  \label{fig:wordcloud_category2_lgbtq}
\end{subfigure}%
\begin{subfigure}{.5\columnwidth}
  \centering
  \includegraphics[width=\columnwidth]{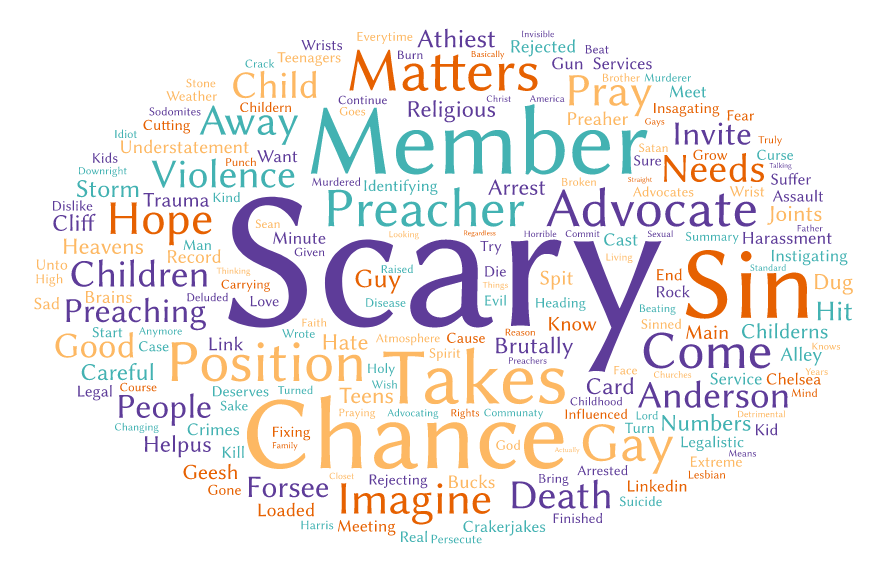}
  \caption{Warning of consequences}
  \label{fig:wordcloud_category3_lgbtq}
\end{subfigure}%
\begin{subfigure}{.5\columnwidth}
  \centering
  \includegraphics[width=\columnwidth]{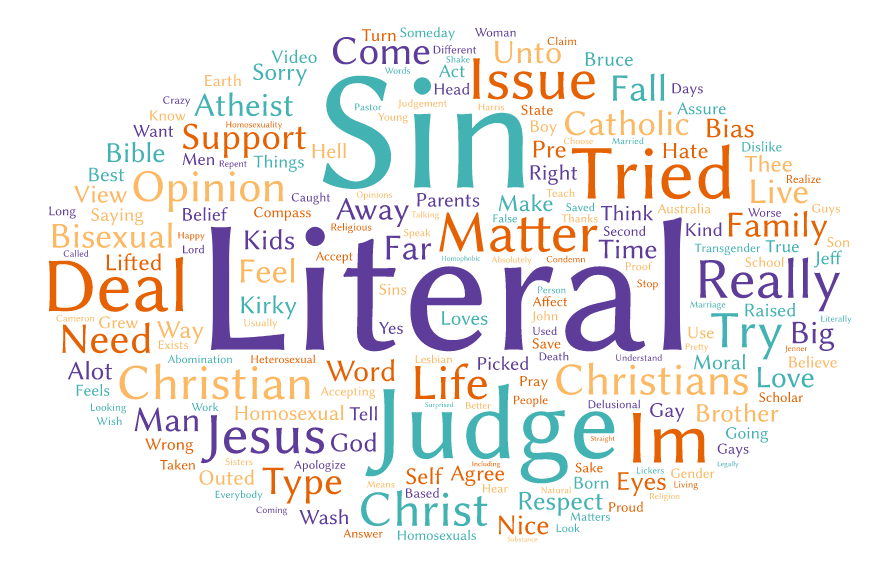}
  \caption{Affiliation}
  \label{fig:wordcloud_category4_lgbtq}
\end{subfigure}%

\begin{subfigure}{.5\columnwidth}
  \centering
  \includegraphics[width=\columnwidth]{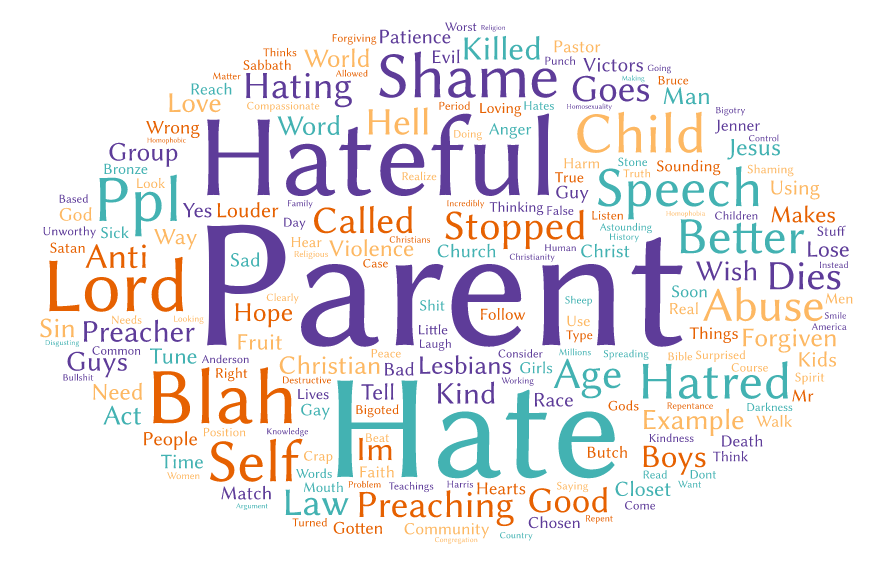}
  \caption{Denouncing speech}
  \label{fig:wordcloud_category5_lgbtq}
\end{subfigure}
\begin{subfigure}{.5\columnwidth}
  \centering
  \includegraphics[width=\columnwidth]{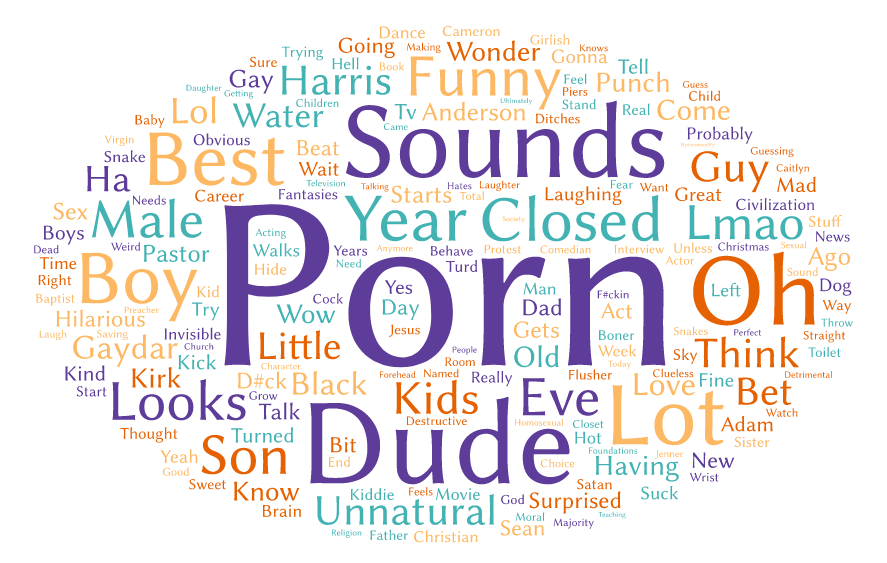}
  \caption{Humor and sarcasm}
  \label{fig:wordcloud_category6_lgbtq}
\end{subfigure}%
\begin{subfigure}{.5\columnwidth}
  \centering
  \includegraphics[width=\columnwidth]{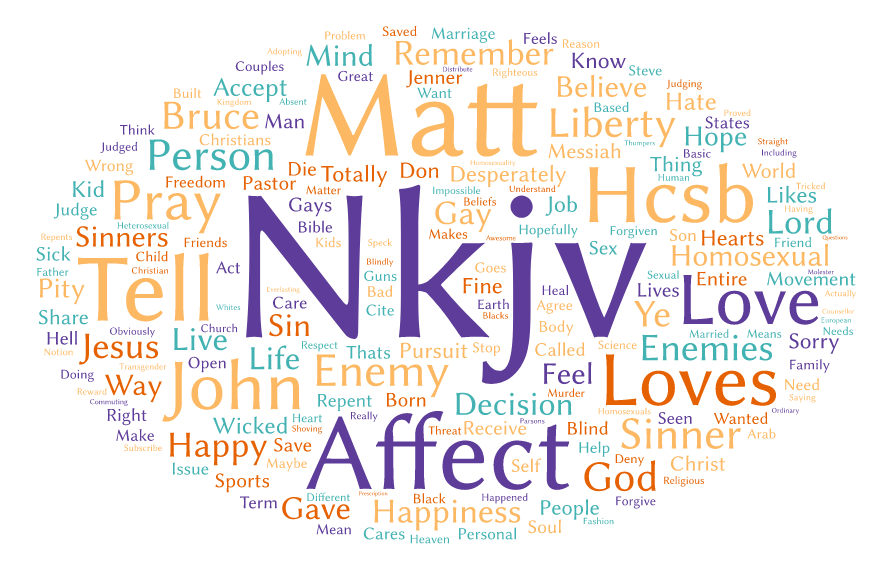}
  \caption{Positive tone}
  \label{fig:wordcloud_category7_lgbtq}
\end{subfigure}%
\begin{subfigure}{.5\columnwidth}
  \centering
  \includegraphics[width=\columnwidth]{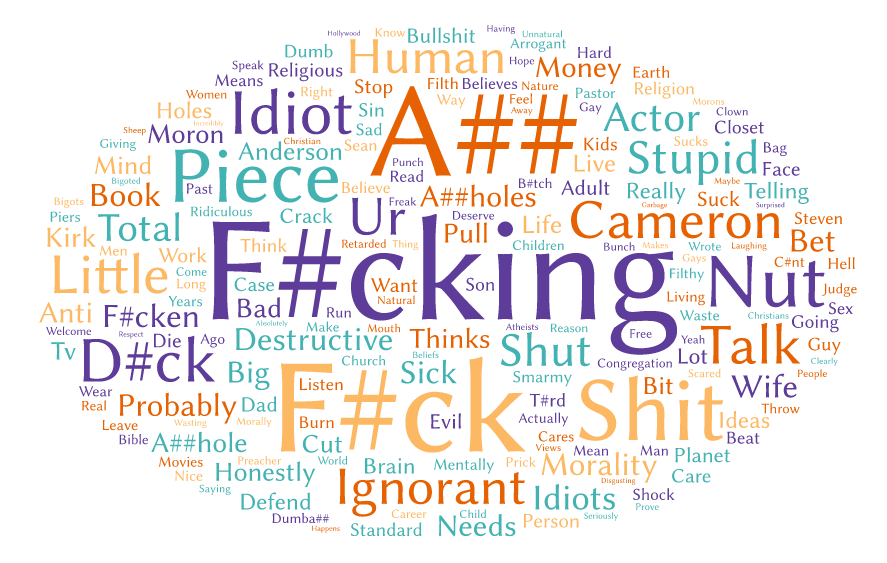}
  \caption{Hostile language}
  \label{fig:wordcloud_category8_lgbtq}
\end{subfigure}%
\caption{Word clouds for the different types of counterspeech used by the counterspeaker for hate speech against \textbf{LGBT}.}
\label{fig:wordcloud_counterspeech_type_lgbtq}

\end{figure*}

In Figure~\ref{fig:wordcloud_counterspeech_type_lgbtq}, we plot the word cloud for the different types of counterspeech employed for the LGBT community. Observe the presence of words such as `Hypocrite' in Figure~\ref{fig:wordcloud_category2_lgbtq}, `Arrest', `Die', `Meet' in Figure~\ref{fig:wordcloud_category3_lgbtq}, `Im', `Christian', `Support' in Figure~\ref{fig:wordcloud_category4_lgbtq}, `Hatred', `Bigot' in Figure~\ref{fig:wordcloud_category5_lgbtq}, `Lmao', `Funny' in Figure~\ref{fig:wordcloud_category6_lgbtq}, `Happiness', `Love' in Figure~\ref{fig:wordcloud_category7_lgbtq}. 

\begin{figure*}[!ht]
\begin{subfigure}{.5\columnwidth}
  \centering
  \includegraphics[width=\columnwidth]{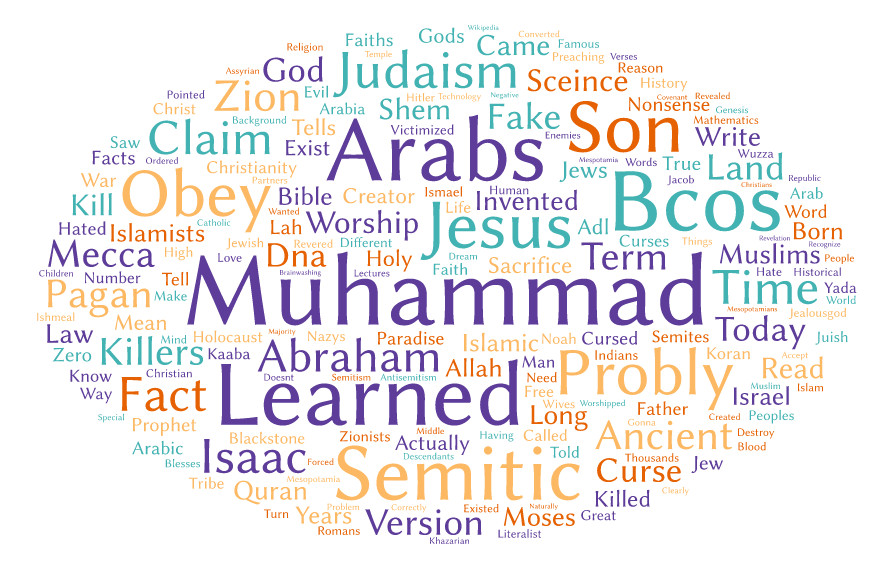}
  \caption{Presenting facts}
  \label{fig:wordcloud_category1_jew}
\end{subfigure}
\begin{subfigure}{.5\columnwidth}
  \centering
  \includegraphics[width=\columnwidth]{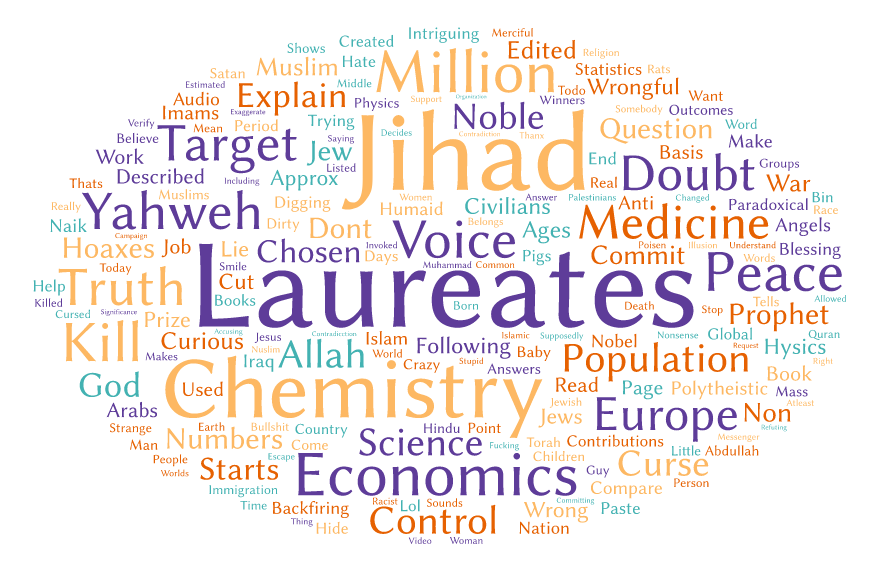}
  \caption{Pointing out hypocrisy}
  \label{fig:wordcloud_category2_jew}
\end{subfigure}%
\begin{subfigure}{.5\columnwidth}
  \centering
  \includegraphics[width=\columnwidth]{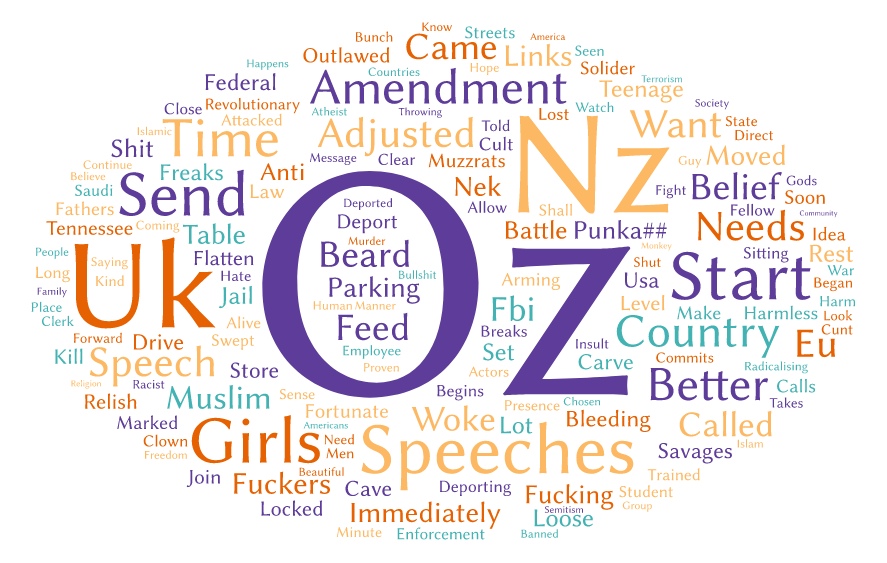}
  \caption{Warning of consequences}
  \label{fig:wordcloud_category3_jew}
\end{subfigure}%
\begin{subfigure}{.5\columnwidth}
  \centering
  \includegraphics[width=\columnwidth]{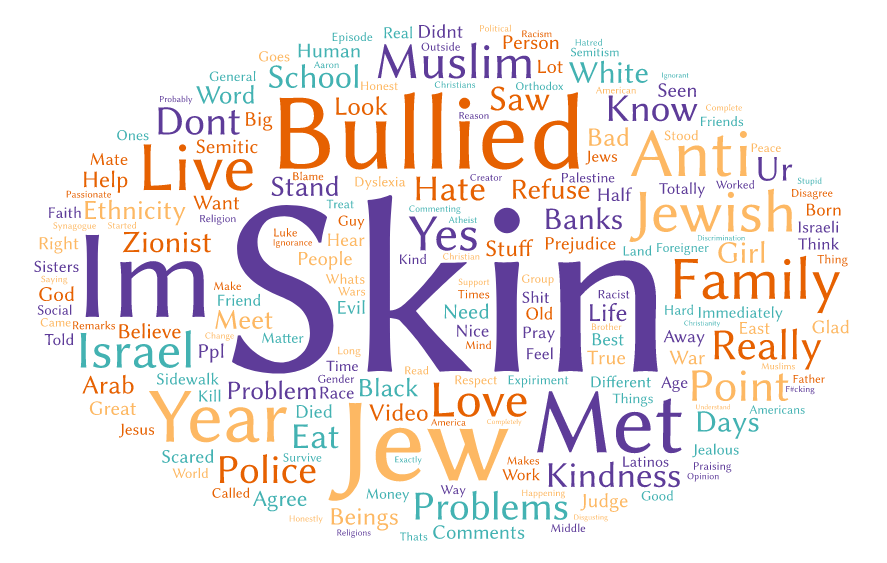}
  \caption{Affiliation}
  \label{fig:wordcloud_category4_jew}
\end{subfigure}%

\begin{subfigure}{.5\columnwidth}
  \centering
  \includegraphics[width=\columnwidth]{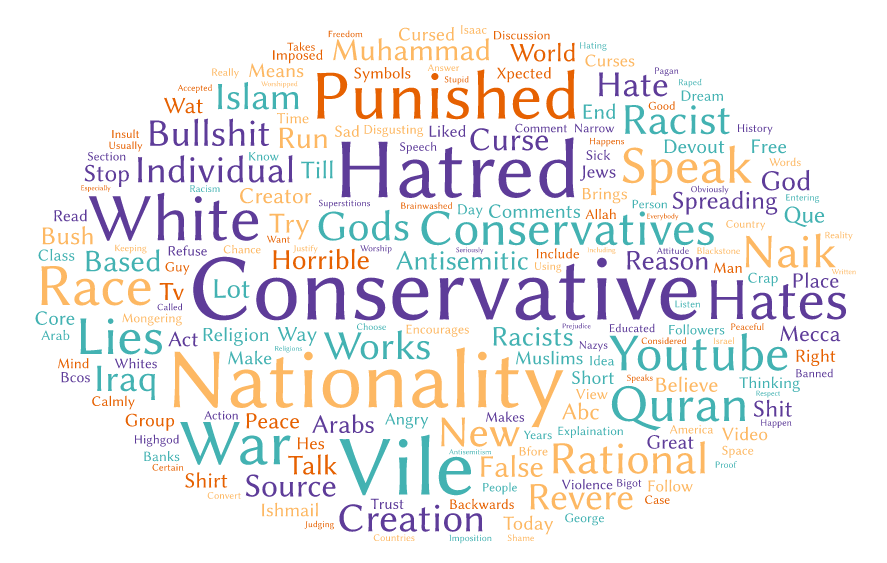}
  \caption{Denouncing speech}
  \label{fig:wordcloud_category5_jew}
\end{subfigure}
\begin{subfigure}{.5\columnwidth}
  \centering
  \includegraphics[width=\columnwidth]{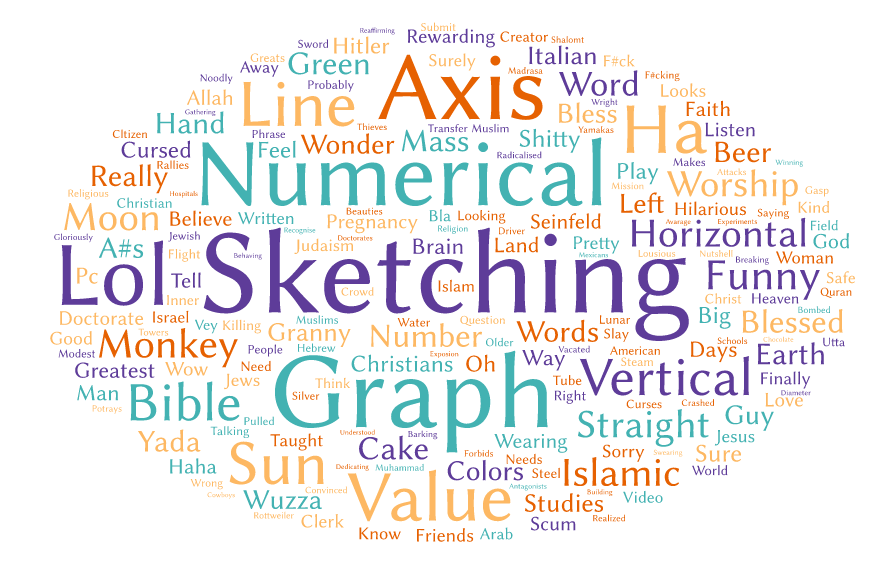}
  \caption{Humor and sarcasm}
  \label{fig:wordcloud_category6_jew}
\end{subfigure}%
\begin{subfigure}{.5\columnwidth}
  \centering
  \includegraphics[width=\columnwidth]{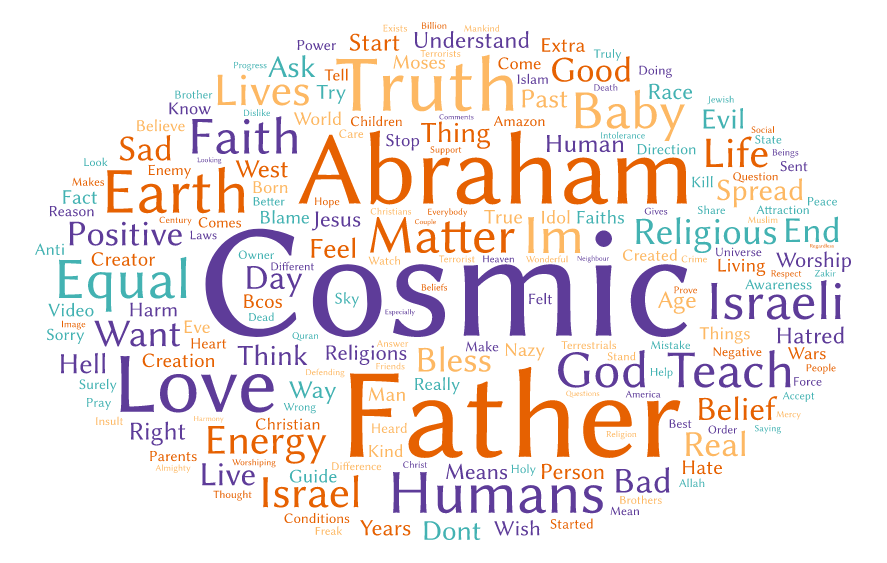}
  \caption{Positive tone}
  \label{fig:wordcloud_category7_jew}
\end{subfigure}%
\begin{subfigure}{.5\columnwidth}
  \centering
  \includegraphics[width=\columnwidth]{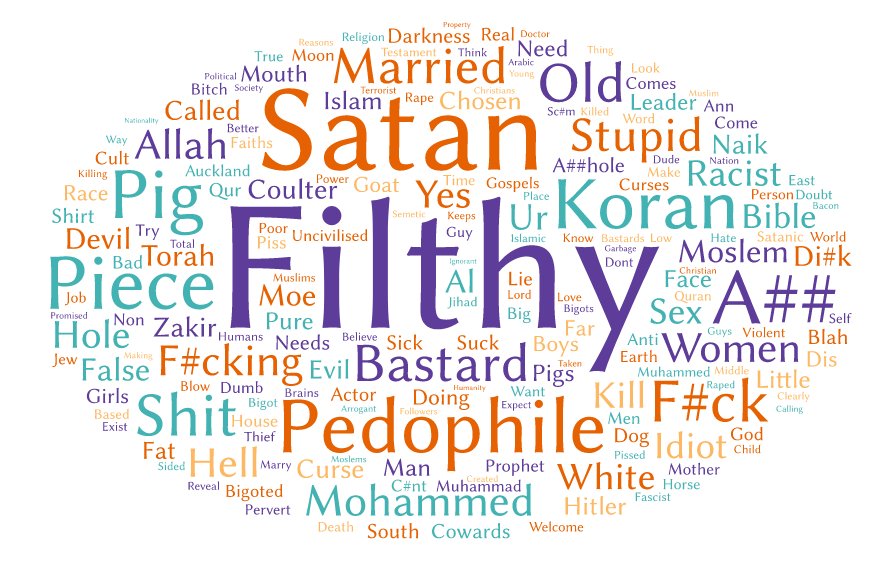}
  \caption{Hostile language}
  \label{fig:wordcloud_category8_jew}
\end{subfigure}%
\caption{Word clouds for the different types of counterspeech used by the counterspeaker for hate speech against \textbf{Jews}.}
\label{fig:wordcloud_counterspeech_type_jew}

\end{figure*}

In Figure~\ref{fig:wordcloud_counterspeech_type_jew}, we plot the word cloud for the different types of counterspeech employed for the Jews community. Observe the presence of words such as `Paradoxical' in Figure~\ref{fig:wordcloud_category2_jew}, `Deporting', `Enforcement', `Jail', and `Fbi' in Figure~\ref{fig:wordcloud_category3_jew}, `Im', `Ur', `Love' in Figure~\ref{fig:wordcloud_category4_jew}, `Hatred', `Racist' in Figure~\ref{fig:wordcloud_category5_jew}, `Lol', `Funny' in Figure~\ref{fig:wordcloud_category6_jew}, `Bless', `Love' in Figure~\ref{fig:wordcloud_category7_jew}.

\subsection{Psycholinguistic analysis}
The language that online users choose, provides important psychological cues to their thought processes, emotional states, intentions, and motivations~\cite{tausczik2010psychological}. The LIWC tool\footnote{LIWC : \url{http://liwc.wpengine.com/}} helps in understanding several psycholinguistic properties used in the text. In order to understand the psycholinguistic differences, we apply LIWC (i.e., the fraction of words in different linguistic and cognitive dimensions identified by the LIWC tool) on both counter and non-counter comments. Finally, we look for statistically significant differences between these two groups with respect to the above analysis. We run Mann–Whitney U test~\cite{mann1947test} and report the significantly different categories in Table~\ref{tab:liwc_analysis}.

We observe several LIWC categories that show significant differences between counter and non-counter comments. The `spoken' category of LIWC (`assent' and `non-fluencies') is more pronounced in non-counterspeech, whereas `affective processes' (`anxiety', `anger', `sadness', `negative emotion' and `affect') are more strong in counterspeech. `Personal concern' (`religion', `achievement', `work', `leisure', and `money') is more pronounced in non-counter comments. The `biological processes' (`body', `health', `sexual'), on the other hand, seems to be more dominant in the language of the counterspeakers.

\begin{table}[!b]
	\centering
	\resizebox{\linewidth}{!}{\begin{tabular}{| p{3cm} |p{2.2cm} | p{2cm} |p{2.2cm}|p{1.5cm}|} 
			\hline
			Dimension &Category & Counter (mean) & Non-counter (mean) & Significance Level\\ \hline
            
\multirow{ 6}{*}{Personal concerns}&Achiev & \cellcolor{red!20}0.334 & \cellcolor{green}\textbf{0.383}&* \\
									&Work & \cellcolor{red!20}0.316 & \cellcolor{green}\textbf{0.397}&**\\
									&Leisure & \cellcolor{red!20}0.179 & \cellcolor{green}\textbf{0.251}&*\\
									&Home & \cellcolor{green}\textbf{0.057} & \cellcolor{red!20}0.046&***\\
									&Money & \cellcolor{red!20}0.123 & \cellcolor{green}\textbf{0.152}&**\\
									&Relig & \cellcolor{red!20}1.148 & \cellcolor{green}\textbf{1.362}&***\\ \hline

\multirow{ 3}{*}{Spoken categories}&Assent & \cellcolor{red!20}0.080 & \cellcolor{green}\textbf{0.095}&** \\
                        &Nonflu & \cellcolor{red!20}0.021 & \cellcolor{green}\textbf{0.031}&*** \\ 
                        &Filler & \cellcolor{green}\textbf{0.162} & \cellcolor{red!20}0.136&*** \\ 
						                  
\hline

\multirow{ 3}{*}{Biological processes}&Body& \cellcolor{green}\textbf{0.263} & \cellcolor{red!20}0.175&*** \\
									 &Health& \cellcolor{green}\textbf{0.131} & \cellcolor{red!20}0.108&*** \\
									 &Sexual & \cellcolor{green}\textbf{0.461} & \cellcolor{red!20}0.357&*** \\
                         
\hline     

\multirow{ 1}{*}{Perceptual processes}&See& \cellcolor{red!20}0.382 & \cellcolor{green}\textbf{0.391}&* \\
                                    &Hear& \cellcolor{red!20}0.259 & \cellcolor{green}\textbf{0.306}&*** \\
                                    &Feel& \cellcolor{green}\textbf{0.084} & \cellcolor{red!20}0.078&*** \\

\hline

\multirow{ 3}{*}{Cognitive processes}&Insight & \cellcolor{green}\textbf{0.660} & \cellcolor{red!20}0.586&*** \\
									&Discrep & \cellcolor{green}\textbf{0.626} & \cellcolor{red!20}0.586&*** \\
                                    &Certain& \cellcolor{red!20}0.551 & \cellcolor{green}\textbf{0.655}&*** \\
                                    &Incl& \cellcolor{red!20}1.400 & \cellcolor{green}\textbf{1.417}&* \\
                                    &Excl& \cellcolor{red!20}0.976 & \cellcolor{green}\textbf{1.067}&** \\
                                    
\hline

\multirow{ 6}{*}{Affective processes}&Anx & \cellcolor{green}\textbf{0.121} & \cellcolor{red!20}0.086 &***\\
									&Negemo & \cellcolor{green}\textbf{1.429} & \cellcolor{red!20}1.089 &***\\
                                    &Posemo & \cellcolor{red!20}1.066 & \cellcolor{green}\textbf{1.149}&*** \\
                                    &Affect& \cellcolor{green}\textbf{2.488} & \cellcolor{red!20}2.217 &***\\
                                    &Anger & \cellcolor{green}\textbf{0.937} & \cellcolor{red!20}0.654 &***\\
                                    &Sad & \cellcolor{green}\textbf{0.093} & \cellcolor{red!20}0.079 &**\\

\hline

\multirow{ 1}{*}{Social processes}&Humans& \cellcolor{green}\textbf{0.759} & \cellcolor{red!20}0.621 &***\\
                                    &Family & \cellcolor{green}\textbf{0.113} & \cellcolor{red!20}0.105 &***\\
                                    &Friends & \cellcolor{green}\textbf{0.042} & \cellcolor{red!20}0.033 &***\\

\hline

\multirow{13}{*}{Linguistic processes}&Funct& \cellcolor{red!20}17.013 & \cellcolor{green}\textbf{17.811} &***\\
									&Swear & \cellcolor{green}\textbf{0.353} & \cellcolor{red!20}0.164 &***\\
                                    &I& \cellcolor{green}\textbf{0.658} & \cellcolor{red!20}0.543 &***\\
                                    &Ipron & \cellcolor{green}\textbf{2.006} & \cellcolor{red!20}1.951 &***\\
                                    &Negate & \cellcolor{red!20}0.779 & \cellcolor{green}\textbf{0.859} &***\\
                                    &Past & \cellcolor{red!20}0.746 & \cellcolor{green}\textbf{0.941} &***\\
                                    &Present & \cellcolor{green}\textbf{3.389} & \cellcolor{red!20}3.301&*** \\
                                    &Pronoun& \cellcolor{green}4.281 & \cellcolor{red!20}\textbf{4.161}&*** \\
                                    &They & \cellcolor{red!20}0.441 & \cellcolor{green}\textbf{0.566} &***\\
                                  	&Verbs & \cellcolor{red!20}4.888 & \cellcolor{green}\textbf{4.957} &***\\
                                  	&You & \cellcolor{red!20}0.517 & \cellcolor{green}\textbf{0.541}&* \\
                                    &SheHe & \cellcolor{green}\textbf{0.683} & \cellcolor{red!20}0.578&*** \\

			\hline
\end{tabular}}
\caption{LIWC analysis of the counter and non-counter comments. Only those LIWC categories are shown which are statistically significant: $p < 0.05$ (*), $p < 0.01$ (**), $p < 0.001$ (***). Note that each LIWC category is either dense in green cells (red cells) for the counter (non-counter) comments or for the non-counter (counter) comments.}
~\label{tab:liwc_analysis}

\end{table}

\section{Classification model}
We consider three classification tasks that naturally manifest in this problem context. The first task is a binary classification problem in which we present the system with a comment and the task is to predict whether the comment is a counterspeech or non-counterspeech. The second one is a multi-label classification task in which we present the system with a known counterspeech comment and the task is to predict all the types of counterspeech present in the comment. The third task is similar to first, except that it is cross-community, i.e., while the training data is drawn from two of the three communities, the test data is drawn from the remaining community. 

\noindent\textbf{Preprocessing:} Before the classification, we preprocess all the data by eliminating URLs, numerals, stopwords and punctuations\footnote{We did not observe any significant change in the scores by including the stopwords and punctuations.}. The text is then lower cased, tokenized and used as input for the classification pipeline.

\noindent\textbf{Features:} For the task of classification we use \textit{tf-idf} vectors (TF-IDF), \textit{bag of words} vectors (BoWV) and \textit{sentence} vectors (SV). The BoWV approach uses the average of the GloVe~\cite{pennington2014glove} word embeddings to represent a sentence. We set the size of the vector embeddings to 300. The sentence vector is generated using a Universal Sentence Encoder~\cite{cer2018universal} which outputs a 512 dimensional vector representation of the text. Recent works~\cite{conneau2017supervised} have shown better performance using pre-trained sentence level embeddings as compared to word level embeddings.

\noindent\textbf{Choice of classifiers:}
We experiment with multiple classifiers such as Gaussian Naive Bayes (GNB), Random Forest (RF), Logistic Regression (LR), SVMs, XGBoost (XGB), CatBoost (CB)~\cite{dorogush2017catboost}, Decision Tree (DT), and neural models such as Multi-layer Perceptron (MLP), LSTM.

\subsection{Counterspeech classification}
In this task, a binary classifier is built to predict if the given input text is a counterspeech or non-counterspeech. We perform stratified 10-fold cross validation on the dataset. 
The whole training set is partitioned into 10 folds, one is set apart for testing, and the remaining nine are used to train the model and evaluate it on the test fold.

The process is repeated 10 times until each fold is used for testing exactly once. We use a held-out validation set to fine tune the parameters of the classifier. The results are computed from the 10 tests and the means and standard deviations of different evaluation measures are reported in Table~\ref{tab:prediction_task1_results}. We use accuracy along with weighted precision, recall, and F1-score as the evaluation measure. 

Among the different features, we observe that sentence vectors seem to be performing much better than BoWV and TF-IDF in most of the cases. We got the best performance using XGBoost along with SV+TF-IDF+BoWV as features. Classifiers such as MLP and CatBoost also performed comparably well. Our best performing model achieves an accuracy of 71.6\%. In the table we show some of the best results obtained using different classifier choices. The results from all the different classifiers and the different feature types are available\footref{dataset_link}.

\begin{table}[!h]
	\centering
\resizebox{\linewidth}{!}{\begin{tabular}{| l | l | l |l| l|} 
		\hline
		Method  & Precision & Recall & F1-Score & Accuracy \\
		\hline \hline

        \rowcolor{green} XGB+SV+TF-IDF+BOWV &	0.716(+/-0.038) &	0.715(+/-0.039) &	0.715(+/-0.04) &	0.716(+/-0.038) \\
        \rowcolor{green!30} MLP+SV+TF-IDF &	0.714(+/-0.031) &	0.713(+/-0.033) &	0.713(+/-0.033) &	0.714(+/-0.032)  \\
        \rowcolor{green!10} CB+SV+TF-IDF+BOWV & 0.708(+/-0.04) & 0.706(+/-0.043) & 0.705(+/-0.043) & 0.707(+/-0.042) \\
         RF+SV+TF-IDF+BOWV	& 0.697(+/-0.043) &	0.693(+/-0.045) &	0.692(+/-0.046) &	0.695(+/-0.044)\\
        SVC+SV+TF-IDF+BOWV &	0.693(+/-0.029) &	0.691(+/-0.03) &	0.691(+/-0.03) &	0.692(+/-0.029) \\
        
		\hline
		
	\end{tabular}}
	\caption{Classification scores for the task of predicting if the given comment is counterspeech or non-counterspeech. The values reported are the means and standard deviations over 10-folds for each of the evaluation metric.}
	~\label{tab:prediction_task1_results}

\end{table}

\subsection{Counterspeech type classification}
Here, we build models for a multi-label classification task in which the input to the classifier is a counter comment and the output are the types of counterspeech present in the comment. As a baseline, we use $General_B$~\cite{metz2012estimation} which predicts the top most frequent labels of the dataset based on the cardinality\footnote{Cardinality represents the average size of the multi-labels in the dataset, which is 1.32 in our case. We follow the same procedure as the authors and take the closest integer value of the cardinality as the cardinality of the dataset (which will be 1).} of the dataset. For our dataset, only the most frequent label `Hostile language' was predicted to be relevant.

The performance of a multi-label classifier should be always assessed by means of several evaluation metrics~\cite{madjarov2012extensive}. In this paper, the multi-label classifiers are evaluated using five measures: accuracy, precision, recall, F1-score and hamming loss~\cite{godbole2004discriminative,schapire2000boostexter}. 

Accuracy is defined as the proportion of predicted \textit{correct} labels to the \textit{total} number of labels, averaged over all instances.

\begin{equation}
Accuracy = \frac{1}{\mid D \mid} \displaystyle\sum\limits_{i=1}^{\mid D \mid} \frac{\mid Y_i \cap Z_i \mid }{\mid Y_i \cup Z_i \mid } \label{Precision1}
\end{equation}

Precision is defined as the proportion of predicted \textit{correct} labels to the total number of \textit{actual} labels, averaged over all instances

\begin{equation}
Precision = \frac{1}{\mid D \mid} \displaystyle\sum\limits_{i=1}^{\mid D \mid} \frac{\mid Y_i \cap Z_i \mid }{\mid Z_i \mid } \label{Precision2}
\end{equation}

Recall is defined as the proportion of predicted \textit{correct} labels to the total number of \textit{predicted} labels, averaged over all instances

\begin{equation}
Recall = \frac{1}{\mid D \mid } \displaystyle\sum\limits_{i=1}^{\mid D \mid } \frac{\mid Y_i \cap Z_i \mid}{\mid Y_i \mid} \label{Recall}
\end{equation}

F1-score is defined simply as the harmonic mean of precision and recall.
\begin{equation}
\textit{F1-score} = 2 * \frac{Precision * Recall}{Precision + Recall} \label{F1}
\end{equation}

Hamming loss  is equal to 1 over
$|D|$ (number  of  multi-label  samples),  multiplied  by  the  sum  of  the symmetric  differences between  the  predictions ($Z_i$)  and  the  true  labels ($Y_i$),  divided  by  the  number  of labels (L),  giving

\begin{equation}\small
Hamming\textrm{ }Loss = \frac{1}{|D|} \displaystyle\sum\limits_{i=1}^{|D|} \frac{|Y_i \Delta Z_i|}{|L|} . \label{HL}
\end{equation}

All these performance measures have values in the interval $[0 ... 1]$. For Hamming loss, the smaller the value, the better the multi-label classifier performance is, while for the other measures, the greater values indicate better performance.

For the multi-label classification, we perform multi-label stratified\footnote{\url{https://github.com/trent-b/iterative-stratification}} 10-fold cross validation ~\cite{sechidis2011stratification} on the dataset. The whole  training  set  is  partitioned  into  10  folds,  one  is  set apart  for  testing,  and  the  remaining  nine  are  used  to  train the model and evaluate it on the test fold. The process is repeated 10 times until each fold is used for testing exactly once. We use a held-out validation set to fine tune the parameters of the classifier. The results are computed from the 10 tests and the means and standard deviations of different evaluation measures are reported in Table~\ref{tab:prediction_task2_results}. We obtain the best performance on using XGBoost with SV+TF-IDF+BoWV as the feature set.The results from all the different classifiers and the different feature types are available\footref{dataset_link}.

\begin{table}[!htbp]
	
\resizebox{\linewidth}{!}{\begin{tabular}{| p{2.0cm}  |  l|l|l|l|p{2.3cm}|} 
		\hline
		Method  & Accuracy & Precision & Recall & F1-Score & Hamming Loss \\
		\hline
        $General_B$	     	 & 0.322 & 0.397 & 0.322 & 0.356 & \cellcolor{green!30}0.191 \\

        XGB+SV+TF-IDF+BOWV &	\cellcolor{green}0.472(+/-0.012) &	\cellcolor{green}0.509(+/-0.012) &	\cellcolor{green!30}0.733(+/-0.011) &	\cellcolor{green}0.601(+/-0.011) &	0.212(+/-0.015) \\
        MLP+SV+TF-IDF &	\cellcolor{green!10}0.44(+/-0.014) &	\cellcolor{green!30}0.504(+/-0.013) &	0.527(+/-0.021) &	\cellcolor{green!10}0.515(+/-0.015) &	0.295(+/-0.018) \\
        LR+SV+TF-IDF &	\cellcolor{green!30}0.469(+/-0.014) &	\cellcolor{green!10}0.5(+/-0.014) &	\cellcolor{green}0.734(+/-0.02) &	\cellcolor{green!30}0.595(+/-0.015) &	0.212(+/-0.015) \\
        GNB+SV &	0.339(+/-0.014) &	0.357(+/-0.016) &	\cellcolor{green!10}0.71(+/-0.02) &	0.475(+/-0.018) &	\cellcolor{green}0.072(+/-0.012) \\
        DT+SV+TF-IDF &	0.301(+/-0.015) &	0.356(+/-0.02) &	0.361(+/-0.02) &	0.358(+/-0.019) &	\cellcolor{green!10}0.193(+/-0.012) \\

		\hline
	\end{tabular}}
	\caption{Classification scores for the task of multi-label classification of the types of counterspeech. The values reported are the means and standard deviations over 10-folds for each of the evaluation metric.}
	~\label{tab:prediction_task2_results}
\end{table}

To get a better understanding of how the classifier is performing for each counterspeech type, we look into the label-wise performance as illustrated in Table~\ref{tab:label_wise_performance}. These label-wise results are obtained using the best-classifier (XGBoost with SV+TF-IDF+BoWV) we obtained in the previous step. We observe that the classifier is able to perform good for certain classes such as `Hostile' and `Affiliation', while it performs poorly for other types such as `Warning of offline/online consequences' and `Pointing out Hypocrisy/contradiction'.

\begin{table}[!htbp]
	\centering
\resizebox{0.95\columnwidth}{!}{\begin{tabular}{|l|l|l|l|} 
		\hline
		Counterspeech Type  & Precision & Recall & F1-Score \\
		\hline
        	Presenting facts & 0.443(+/-0.036) & 0.648(+/-0.049) & 0.526(+/-0.040) \\
			Pointing out hypocrisy & 0.353(+/-0.026) & 0.630(+/-0.043) & 0.452(+/-0.031) \\
			Warning of  consequences & 0.434(+/-0.057) & 0.470(+/-0.087) & 0.450(+/-0.066) \\
			Affiliation & 0.567(+/-0.052) & 0.605(+/-0.084) & 0.582(+/-0.052) \\
			Denouncing hateful speech & 0.413(+/-0.023) & 0.694(+/-0.041) & 0.518(+/-0.025) \\
			Humor & 0.462(+/-0.036) & 0.661(+/-0.042) & 0.543(+/-0.036) \\
			Positive tone & 0.430(+/-0.050) & 0.598(+/-0.046) & 0.500(+/-0.047) \\
			Hostile & 0.547(+/-0.016) & 0.919(+/-0.017) & 0.686(+/-0.015) \\

		\hline
	\end{tabular}}
	\caption{The performance scores of each type of counterspeech as given on using the XGB+SV+TF-IDF+BoWV. The values reported are the means and standard deviations over 10-folds for each of the evaluation metric.}
	~\label{tab:label_wise_performance}

\end{table}

\subsection{Cross-community classification}
In this section, we build models that draw the training data points  from  two communities  to  predict  the  labels  for the test data drawn from the third community. In this task, a binary classifier is built to predict if the given input text is a counterspeech or non-counterspeech.  Since this task is the same as first task (counterspeech classification), we use the best model (XGBoost+SV+TF-IDF+BoWV) that we obtained from the first task. The training consists of data points from two communities and the test set will be the third community. The process is repeated until each community is used for testing exactly once. Note that this application is motivated by the fact that in the context of the current problem there might exist communities for which in-community training instances are scarce and therefore the only way to perform the classification is to resort to the training instances available for other communities (see~\cite{Rudra:2015} for a similar approach). For evaluation, we report accuracy, weighted precision, recall and F1-score. Table~\ref{tab:prediction_task3_results} shows the results of this task. The models are able to produce comparable results even while they are trained using instances from a different community. This is an extremely desirable feature to avoid requirement of fresh annotations every time the model is used for a new  (and so far unseen) community.

\begin{table}[!htbp]
	\centering
\resizebox{0.95\columnwidth}{!}{\begin{tabular}{| l| l| l| l| l | l |l| l|} 
		\hline
		Train  & Test & Precision & Recall & F1-Score & Accuracy\\
		\hline
		Blacks+Jews & LGBT & 0.652 & 0.655 & 0.653 & 0.644 \\ 
		Jews+LGBT & Blacks & 0.617 & 0.616 & 0.617 & 0.616 \\
		LGBT+Blacks & Jews & 0.621 & 0.628 & 0.624 & 0.620 \\
		
		\hline
		
	\end{tabular}}
	\caption{Classification scores for the task of predicting if the given comment is counterspeech or non-counterspeech in one community using the training instances from the other two communities.}
	~\label{tab:prediction_task3_results}

\end{table}

\begin{table}[!tbp]

	\centering
\resizebox{0.95\columnwidth}{!}{\begin{tabular}{| l| l| l| l| l| l| l| } 
		\hline
		&\multicolumn{2}{|c|}{Jews} & \multicolumn{2}{|c|}{LGBT} &  \multicolumn{2}{|c|}{Blacks} \\
		\hline
		Type of Counterspeech  & \#R  & \%A & \#R  & \%A & \#R  & \%A \\
		\hline
    Presenting facts & 112 & 28.57 & 49 & 46.94 & 22 & 36.36 \\
    Pointing out hypocrisy or contradictions & 114 & 42.98 & 68 & 41.18 & 33 & 39.39 \\
    Warning of offline or online consequences & 45 & 71.11 & 30 & 26.67 & 55 & 34.54 \\
    Affiliation & 121 & 51.24 & 76 & 40.79 & 86 & 1.16 \\
    Denouncing hateful or dangerous speech & 124 & 44.35 &85 & 43.53 & 120 & 19.17 \\
    Humor & 62 & 46.77 & 101 & 65.35 & 49 & 28.57 \\
    Positive tone & 156 & 42.31 & 54 & 38.89 & 165 & 6.06  \\
    Hostile & 87 & 35.63 & 105 & 42.86 & 66  & 48.48 \\
		\hline
		
	\end{tabular}}%
	\caption{Percentage of replies that agree with the opinion of the counter speaker.`\#R' represents the number of replies that were tagged and `\%A' represents the percentage of replies that agree with the counterspeakers comment.}
	~\label{tab:counterspeech_agreement}

\end{table}

\section{Discussion}

We observe that non-counterspeech consists mainly of comments that agree with the main content in the video or hate speech toward the target community itself. These vary depending on the community involved. In case of {\sl Jews}, we find that majority of the comments claimed that the Jews are controlling the economy and are responsible for the destruction of their society. Many of the non-counterspeech also included holocaust denial~\cite{gerstenfeld2003hate}. In case of {\sl Blacks}, we find that the majority of non-counterspeech were hate speech in the form of racist remarks such as ni**ers, slavery etc. In case of {\sl LGBT}, we observe that the majority of non-counterspeech are linked to religious groups claiming that it is unnatural and forbidden in their religion.

Not all types of counterspeech are equally effective~\cite{susan2016successfullcounter}. To understand the nature of replies received by each type of counterspeech comments, we investigated into the people's reaction to the counterspeech. This would tell us how the community views these statements provided by the counterspeakers. For each target community and for each type of counterspeech, we randomly select 10 counterspeech comments that have at least two replies. Next, we ask the annotators to check if the response to the counterspeech comment is in agreement with the counterspeaker's opinion or against it. We report the results in Table~\ref{tab:counterspeech_agreement}.  As observed from the table, the community acceptance (as observed by \% agreement) of the type of counterspeech varies.  In Jew community, counterspeech strategies involving `Warning offline or online consequences' and `Affiliation' seem to be more favoured by the community.  In case of the LGBT community, `Humor' seems to be the most acceptable form of counterspeech by the community while `Warning offline or online consequences' seems to be the least favored tactics. In case of the Blacks community, we observe that counterspeech strategies such as `Affiliation' and `Positive Tone' receives very less community acceptance. We found several cases, in which the replies to such counterspeech were stating that such `Positive Tone' and `Affiliation' will not change the stance of the hate speakers. Although, using `Hostile language' seems to be very prevalent (see Table~\ref{tab:dataset_statistics}), we found that this strategy is not welcomed by even the target community in whose favor these are posted. In many instances, the target community users tend to oppose this form of counterspeech and request the counterspeakers to refrain from using such language of hate.

The counterspeech classifiers can be used in several scenarios. One such promising area, is studying the effectiveness of the types of counterspeech on a larger scale. One could also use such classifiers to generate datasets that could potentially be used to build systems that automatically counter hate messages in online social media. Such a system would be very effective as they could characterize the hate speaker and provide an effective counterspeech based on his/her profile.

\section{Conclusion and future works}

The proliferation of hateful content in online social media is a growing concern. Currently used methods such as blocking or suspension of messages/accounts cause problems to the freedom of speech. Counterspeech is emerging as a very promising option backed by several organizations and NGOs. With no dataset and model available for counterspeech detection, no large scale study can be conducted. In this paper, we took the first step toward creating a dataset of counterspeech against hateful videos in YouTube. We found that counter comments receive more likes than non-counter comments. Further, the psycholinguistic analysis of the comments reveal striking differences between the language choice of counter and non-counter speakers. We found that different communities seem to have different preferences for the selection of counterspeech type. Our models and dataset are placed in the public domain.

There are several directions, which can be taken up as future research. One immediate step is to develop automatic counterspeech detection models for other social media sites like Facebook and Twitter. Another direction could be to study the effectiveness of different types of counterspeech for different communities. A connected research objective could be to investigate how effective the counterspeakers are in changing the mindset of the hate users.

\bibliography{Main}
\bibliographystyle{aaai}

\end{document}